\documentclass{sig-alternate-05-2015}
\newfont{\mycrnotice}{ptmr8t at 7pt}
\newfont{\myconfname}{ptmri8t at 7pt}
\let\crnotice\mycrnotice%
\let\confname\myconfname%

\usepackage{epsfig,balance,cite}
\usepackage[ruled,lined]{algorithm2e}

\begin{document}
\title{Machine to Machine (M2M) Communications in Virtualized Vehicular Ad Hoc Networks}

\numberofauthors{2} 
%
\author{
%
%
\alignauthor
Meng Li\\
       \affaddr{Beijing Advanced Innovation Center for Future Internet Tech.}\\
       \affaddr{College of Electronic Info. and Control Eng.}\\
       \affaddr{Beijing University of Tech., Beijing, P.~R.~China}\\
       \email{limeng0720@emails.bjut.edu.cn}
\and
\alignauthor
F. Richard Yu\\
       \affaddr{Depart. of Systems and Computer Eng.} \\
       \affaddr{Carleton University, Ottawa, ON, Canada}\\
       \email{richard.yu@carleton.ca}
\and
\alignauthor
Pengbo Si, Enchang Sun, Yanhua Zhang\\
       \affaddr{Beijing Advanced Innovation Center for Future Internet Tech.}\\
       \affaddr{College of Electronic Info. and Control Eng.}\\
       \affaddr{Beijing University of Tech., Beijing, P.~R.~China}\\
      \email{\{sipengbo,ecsun,zhangyh\}@bjut.edu.cn}
}

\maketitle

\begin{abstract}
With the growing interest in the use of internet of things (IoT), machine-to-machine (M2M) communications have become an important networking paradigm. In this paper, with recent advances in wireless network virtualization (WNV), we propose a novel framework for M2M communications in vehicular ad-hoc networks (VANETs) with WNV. In the proposed framework, according to different applications and quality of service (QoS) requirements of vehicles, a hypervisor enables the virtualization of the physical vehicular network, which is abstracted and sliced into multiple virtual networks. Moreover, the process of resource blocks (RBs) selection and random access in each virtual vehicular network is formulated as a partially observable Markov decision process (POMDP), which can achieve the maximum reward about transmission capacity. The optimal policy for RBs selection is derived by virtue of a dynamic programming approach. Extensive simulation results with different system parameters are presented to show the performance improvement of the proposed scheme.
\end{abstract}

\category{H.4}{Information Systems Applications}{Miscellaneous}

\keywords{Machine-to-machine (M2M) communications; vehicular ad-hoc networks (VANETs); random access; wireless network virtualization (WNV).}

\section{Introduction}
In recent years, with the rapid development of internet of things (IoT) and wireless network technologies, machine-to-machine (M2M) communications have been attracting lots of attention from industry and academia~\cite{WC15,AK14,NN14,TL14,AB15,MKB14}. European Telecommunications Standards Institute (ETSI) estimated that M2M principles apply particularly well to networks where a large number of machines are used, even up to the 1.5 billion wireless devices in the future~\cite{LL11}. Therefore, M2M communications are taken into account to create a revolution in our future IoT world~\cite{LA14}.\

As the main distinct characteristic from traditional wireless communications, a large number of M2M devices are usually involved in most applications and scenarios based on M2M communications~\cite{RS15}. Meanwhile, as one of the application scenarios, the vehicular networks also have the same feature. Since the number of vehicles has increased considerably every year, the vehicular networks have to be able to sustain an increasing number of vehicles~\cite{ZZ15,BZ11,WY14}. Moreover, for the vehicular networks, the main characteristic is not fixed in size, and the network needs to be scalable to adapt to the continuously and diversely changing network topology due to the nodes' high mobility~\cite{VB15,BG12}. Furthermore, during each time slot, the data transmission with M2M communications in vehicular networks is usually deemed as small-sized, while the frequency of their making data connections is higher than traditional devices since they have specific applications and functions~\cite{WY15}, e.g., Global Positioning System (GPS) in the vehicles. Therefore, how to support more vehicles simultaneously connecting and accessing to the network is an important and inevitable issue~\cite{CY13}.

Several research efforts have investigated M2M communications  and resource allocation for vehicular networks. In~\cite{PL16}, the authors considered a heuristic resource allocation scheme in vehicular networks, which have significantly less computational complexity than the iterative scheme and negligible performance degradation. A cooperative control policy of collection of vehicles operating is proposed by~\cite{GW15} in the vehicular ad-hoc network (VANET) with access-constrained fading channels. In~\cite{JL16}, the authors proposed an optimization policy for joint resource allocation problem  about resource blocks (RBs) and power in moving small cells, with the goal of enhancing the quality of service (QoS), and solved the optimization problem based on iterative resource allocation algorithm (IRAA).\

With recent advances in wireless communications and networks \cite{MYL04, YL01, LYH10, YK07, AK14,XYJL12, ATV12, YWL06_MONET,LY15, WYS10,RS15, GYJ10, BYC12, XYJ12, YTH09, YHT10, LYJ10, YKL06,JL16,YZX11, LYJ15,GYJ11,BY14,LYL09,YYG15,BYY15,BZ11,ZYN12_JSAC,ZYL10,WTY14,BYL11_Online,BY13}, some excellent works have been done in the vehicular networks with M2M communications. However, most existing researches ignore an important issue that the vehicles may fail to access the network if there is no enough radio resource allocated to the random access (RA) process~\cite{ZK15}. In addition, only one class of vehicular networks is considered in most existing works. However, in practical networks, there are multiple classes of vehicles with different applications or services with M2M communications, such as military applications, emergency services, environmental monitoring, safety and collision avoidance for driverless cars, etc.~\cite{KL14}. Obviously, different applications have different QoS requirements, and they should be treated differently.\

In this paper, with recent advances in \emph{wireless network virtualization} (WNV)~\cite{LY15,LYZ15,LY15m,LYJ15}, a novel framework for M2M communications in vehicular networks with WNV is proposed~\cite{HS11}. The vehicular networks' capacity can be improved by M2M communications to support many characteristics of modern vehicles such as cross-platform networking, autonomous monitoring, and visualization of the devices and information~\cite{BG12}. Meantime, using WNV and software-defined networking techniques \cite{sdnopenissue,CYY15,CYL15,YYG15}, a physical vehicular network can be sliced into multiple virtual vehicular networks, then differentiated vehicular network services can be provided based on differentiated QoS requirements~\cite{LYZ15}. Moreover, a decision-theoretic approach to optimize the random access process of M2M communications can be developed. Specifically, RA process in M2M communications is formulated as a partially observable Markov decision process (POMDP). The vehicles can obtain the maximum transmission rate from the belief state in POMDP optimization policy, which encapsulates the history of system state and access decision.\

The rest of this article is organized as follows. System model is presented in Section \ref{sec:Systemmodel}. In Section \ref{sec:Algorithm1}, an optimization policy for the random access process via POMDP formulation is proposed, and the optimal RBs selection scheme will be given. Simulation results are compared and analysed in Section \ref{sec:Simulation}. Finally, we conclude the proposed work in Section \ref{sec:Conclusion} with future works.\

\section{System Model}\label{sec:Systemmodel}
In this section, we firstly describe the system model for the vehicular network with M2M communications. Following the general frameworks of WNV, there are three parts in the system model, which are described as follows.\

\subsection{Physical Resource in Vehicular Network}
\begin{figure}[!t]
\centering
\includegraphics[width=3.4in]{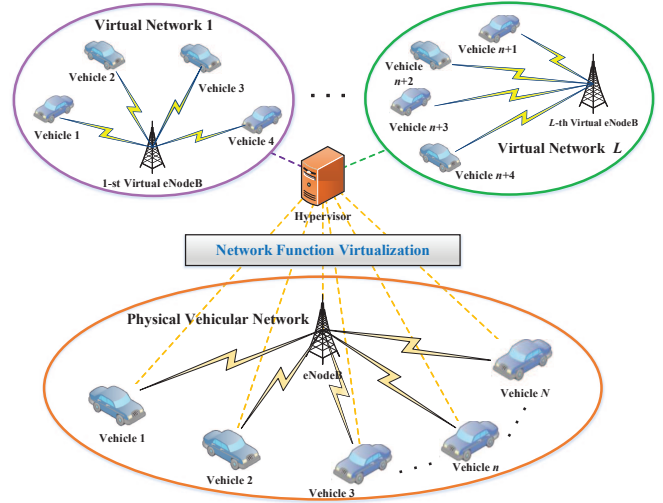}
\caption{The architecture of M2M communications in a vehicular network with WNV.}
\label{fig:model}
\end{figure}

Given the system model in Figure~\ref{fig:model}, a single-cell scenario with $N$ vehicles and one eNodeB is considered in the physical vehicular network~\cite{LR15}. Each vehicle is equipped with a machine-type communication devices (MTCDs).  In the proposed system model, $t_{0}, t_{1},\ldots, t_{k},\ldots, t_{K-1}$ can be denoted the time point that the MTCDs can access the eNodeB, where the total number of time slots is represented as $K$, $1\leq k\leq K-1$. Then, $\delta t_k$ denotes the duration of a time slot, it represents as $t_{k}-t_{k-1}=\delta t_k$. Moreover, one time period includes $K$ time slots, from time point $t_{0}$ to $t_{K-1}$. Meanwhile, RBs will be offered when the MTCDs attempt to access the vehicular network. The number of RBs in random access phase denotes $R$, and $r$ represents the $r$-th RB in the access phase, where $1\leq r\leq R$. The state of each RB in one time slot can be described as idle or busy. The set ${s_r}$ is represented as the state of the $r$-th RB, and ${s_{r}}=\{0,1\}$, where $0$ and $1$ stands for the RB is idle or busy, respectively. In this paper, we assume that the number of busy (or idle) RBs in each time slot satisfies Poisson distribution~\cite{SY13}.\

In the access phase, different transmission rates can be offered by the $r$-th RB after the $n$-th MTCD has accessed. Then, $C_{n,r}(k)$ denotes the transmission rate achieved by the $n$-th MTCD in time slot $\delta t_{k}$, it can be represented as\
\begin{eqnarray}
&& C_{n,r}(k)= \nonumber \\
&& \left\{
\begin{array}{lcl}
B_{n,r} \log_2 \left\{ 1+\frac{P_{r}h_{n,r}}{\sigma^{2}}\right\},\ \ \ \ \ \ \ \ \ \ \ \ \ \ \ \ \ \ \  \text{if}\ s_{r}=0,\\
B_{n,r} \log_2 \left\{1+\frac{P_{r}h_{n,r}}{\sum\limits_{n^{'}\neq n,n^{'}\in N} P_{r}h_{n^{'},r}+\sigma^{2}}\right\}, \text{if}\ s_{r}=1,\\
\end{array}
\right.
\end{eqnarray}
where $B_{n,r}$ denotes the bandwidth offered by the $r$-th RB, $P_{r}$ is the transmit power consumed by the $r$-th RB, $h_{n,r}$ ($h_{n^{'},r}$) represents the channel gain when the $n$-th ($n^{'}$-th) MTCD accesses to the network, which follows Gaussian distribution with zero mean and unit variance, and $\sigma^{2}$ represents the noise power~\cite{AB15}.\

\subsection{Wireless Network Virtualization}
In order to realize WNV, the hypervisor is considered as an important component and needs to be set in the vehicular network. Normally, the hypervisor can be implemented at the physical eNodeB, and it provides functions to connect physical resource and virtual eNodeB~\cite{KM12}. Moreover, the hypervisor takes the responsibility of virtualizing the physical eNodeB into a number of virtual eNodeBs, such as virtual devices and RBs, which can be used by different mobile virtual network operators (MVNOs)~\cite{LY15}. Furthermore, the hypervisor is also used to scheduling the air interface resources for WNV.\

In this paper, the physical vehicular network is abstracted and sliced into multiple virtual networks by the hypervisor according to different application functions or QoS requirements. As a result, it can develop the diverse class of virtual vehicular network, and each virtual network includes MTCDs that have same or similar applications.\

\subsection{Virtual Networks}
As can be seen in Figure~\ref{fig:model}, the physical vehicular network with M2M communications can be sliced into $L$ virtual networks by hypervisor. For the $l$-th virtual network, $N_l (1\leq N_l\leq N)$ vehicles equipped with MTCDs are included in this virtual network, which have the same or similar functions. Meanwhile, in the $l$-th virtual network, RBs can be offered by virtual eNodeB to random access transmission in the initial time slot. $R_l (1\leq R_l\leq R)$ represents the number of RBs used for random access. Besides, the obtained transmission rate in each virtual network can be denoted as $C_{1}, C_{2}, \dots, C_{l}, \dots, C_{L}$, where $C_{1}$ and $C_{L}$ represent the obtained average transmission rate in the highest or lowest level virtual network, respectively.

\section{Optimization Policy of Random Access via POMDP}\label{sec:Algorithm1}
In this section, we formulate a decision-theoretic approach via POMDP to optimize the random access in the vehicular network. After that, each tuple of POMDP is described in detail, followed by the reward and optimization objective. Then, the solution of problem formulation will be given.\

\subsection{POMDP Formulation}
Different from Markov modelling, POMDP has a distinctive characteristic since it emphasizes ``partially observable''. In the proposed scheme, the random access can be formulated as a POMDP optimization problem because RBs states cannot be directly observed by MTCDs~\cite{LM14}. Based on the state of RBs that encapsulates the decision and observation history, the MTCDs will take an effective action in each time slot. In what follows, the random access process of vehicles equipped with MTCDs will be formulated as a POMDP optimization policy and given its solution. For simplicity, we discuss the POMDP formulation in the $l$-th virtual network as an example.\

\emph{1) Action Space} \

At the beginning of time slot $\delta t_k$, according to its current belief state, the $n$-th vehicle equipped with MTCD will access to network and determine which action to select~\cite{ZT07}. Let ${\mathcal{A}}$ be the set of all available actions, and the action taken by this MTCD in time slot $\delta t_k$ is denoted as
\begin{equation}
\begin{aligned}
a(k) \in \{0 (no\ access),{RB_1},{RB_2},\dots,{RB_r},\dots,{RB_{R_{l}}}\}.
\end{aligned}
\end{equation}

In set ${\mathcal{A}}$, $0$ denotes that the MTCD will not access the eNodeB and select sleeping mode, and $RB_r$ represents that the MTCD will select the $r$-th RB to access the network.\

\emph{2) State Space and Transition Probabilities}\

The system state space $\mathcal{S}$ is the set of all RB states in the vehicular network, and the state in time point $t_k$ is represented as $s(k)=[{s_{1}(k)}{s_{2}(k)}\dots{s_{r}(k)}\dots{s_{R_l}(k)}]$, where $s(k)\in\mathcal{S}$\cite{WYH13,ZYL10}. Moreover, the state of the $r$-th RB can be represented as
 \begin{equation}
\begin{aligned}
{s_{r}(k)} \in \{0 (idle),1(busy)\}.
\end{aligned}
\end{equation}

We assumed that each RB state in the proposed scheme is discretized. Then, the transition probability of the $r$-th RB state from state $i$ to state $j$ is represented as
 \begin{equation}\label{stateprobability}
\begin{aligned}
p_{ i,j}=Pr.\{s_r(k+1)=j\mid s_r(k)=i\}.
\end{aligned}
\end{equation}

Considering Poisson distribution, Eq. (\ref{stateprobability}) can be calculated by
\begin{equation}
\begin{aligned}
p_{ i,j}={\frac{(\lambda)^{m}}{m!}}e^{-\lambda},
\end{aligned}
\end{equation}
where $\lambda$ denotes the occurred frequency of the state $j$, and $m$ represents the total number of states varying from $i$ to $j$ in long-term statistics.\

\emph{3) Observation Space}\

As mentioned above, the MTCD needs to observe the RB state because it is difficult to acquire the full knowledge of each RB state~\cite{LM14}. Assume that some MTCDs have selected RBs to access virtual network, it means that part of RBs are in busy state during this time slot. Hence, before accessing the network and making decision, the MTCD needs to observe the RB state. Let $\theta_{r}(k)$ be the observation state of the $r$-th RB in time slot $\delta t_k$, where $1\leq r\leq R_l$. $\theta_{r}(k)$ can be given as
\begin{equation}
\begin{aligned}
{\theta_{r}(k)} \in \{0 (idle),1(busy)\}.
\end{aligned}
\end{equation}
Then in time slot $\delta t_k $, the observation state can be represented as ${\theta(k)}=[\theta_{1}(k)\theta_{2}(k)\dots\theta_{r}(k)\dots\theta_{R_l}(k)]$, where $\theta(k)\in{\theta}$, and the set of all observation states can be denoted as ${\theta}$.

Then, the probability of observation state can be denoted as
 \begin{equation}
\begin{aligned}
b^{a(k)}_{s_r(k+1),\theta_r(k)}=Pr.\{\theta_r(k) \mid s_r(k+1), a(k)\}.
\end{aligned}
\end{equation}
According to the action that be taken by MTCD and the observation state, the conditional probability of observation is calculated as
\begin{eqnarray}
b^{a(k)}_{s_r(k+1),\theta_r(k)}=\left\{
\begin{array}{lll}
\rho, &\text {if}~a(k)=RB_r,~\theta_r(k)=0,\\
1-\rho, &\text {if}~a(k)=RB_r,~\theta_r(k)=1,\\
\upsilon, &\text {if}~a(k)=0,~\theta_r(k)=0,\\
1-\upsilon, &\text {if}~a(k)=0,~\theta_r(k)=1,\\
\end{array}
\right.
\end{eqnarray}
where $\rho$ and $\upsilon$ denote the probabilities of false observation, respectively.

\emph{4) Belief State}\

For POMDP optimization policy, belief state is deemed as an important element. Although the MTCD cannot obtain the RB state directly, it can achieve the state from its action decision and observation history encapsulated by the belief state. In other words, a probability distribution over states, is sufficient statistics for the history, which means that the optimal decision can be made based on the belief state.\

Let $\pi(k)=\{\pi_1^{k}, \pi_2^{k},\dots,\pi_{s_r(k)}^{k},\dots,\pi_{\mathcal{S}}^{k}\}, {s_r(k)\in{\mathcal{S}}}$ be the belief space, where $\pi_{s_r(k)}^{k}\in[0,1]$ denotes the conditional probability that the RB state is in $s_r(k)$ at the beginning of time slot $\delta t_k$ prior to state transition. In order to maintain a belief state, the knowledge of the system dynamics and the state transition probabilities in POMDP are considered as necessary.\

According to~\cite{LM14}, the belief state can be updated easily after each state transition to incorporate additional step information into history. In the proposed scheme, the belief state can be updated by using Bayes' rule at the end of each time slot~\cite{LM14,WTY14,CYY12}~\cite{SS73}. It can be calculated as follows,
\begin{equation}\label{beliefstate}
\pi_{{s_r}(k+1)}^{k+1}=\frac{\sum\limits_{s_r(k)}\pi_{s_r(k)}^{k}p(s_r(k),s_r(k+1))b^{a(k)}_{s_r(k+1),\theta_r(k)}}{\sum\limits_{s_r(k) }\sum\limits_{{s_r(k+1)}}\pi_{s_r(k)}^{k}p(s_r(k),s_r(k+1))b^{a(k)}_{s_r(k+1),\theta_r(k)}}.
\end{equation}

According to Eq. (\ref{beliefstate}), the belief states have ability to capture all the history information at $\delta t_k$. As a result, the past actions and observations can be saved by constantly updating the belief state. It means that it is reasonable to make decisions based on the belief state.\

\emph{5) Reward Function and Objective}\

As discussed above, idle RBs may be offered by eNodeB and the MTCD will choose the proper RB to access in each virtual vehicular network. In this paper, we consider the transmission rate as a reward, the maximum transmission rate offered by RB can be used for performance evaluation. Therefore, for each system state, the corresponding transmission rate in virtual network can be represented as
\begin{equation}
C_{l,n}(k)=\max\limits_{{r}\in[1,R_l]}\{C_{l,n,1}(k),\dots, C_{l,n,r}(k),\dots, C_{l,n,R_l}(k)\},
\end{equation}
where $C_{l,n,r}(k)$ denotes the transmission rate offered by the $r$-th RB in time slot $\delta t_k$.

Then, in the proposed scheme, the optimization objective is to maximize the transmission rate that can be achieved by MTCDs. Hence, the system reward in each time slot is represented as
\begin{eqnarray}
Re_{l,n}(k)=\left\{
\begin{array}{lll}
0, &\text {if there is no accessing},\\
C_{l,n}(k), &\text {otherwise},\\
\end{array}
\right.
\end{eqnarray}
and the long-term total discounted reward $Re_{l,n}$ is given as
\begin{eqnarray}
Re_{l,n}=\sum\limits^{K-1}_{k=0}\beta^{K-k-1}Re_{l,n}(k),
\end{eqnarray}
where $\beta\in[0,1]$ represents the discount factor.\

The optimal policy ${\mathcal{U}}$ in the proposed scheme can be represented as a set of action $a(k)$, $0\leq k\leq K-1$, which maximises the expected long-term total discounted reward $Re_{l,n}$ during a time period. Then, the optimal policy is given as
\begin{eqnarray}
{\mathcal{U}}=\{a(k)\}=\arg\max\limits_{a(k)\in \mathcal{A}}E_{K}[Re_{l,n}].
\end{eqnarray}

\subsection{Solving the POMDP Problem}
In this subsection, an optimal policy for selecting RB by the MTCD can be derived, which maximizes the transmission rate in vehicular networks based on the state of each RB. Meantime, the formulation problem in POMDP can be solved by a dynamic programming methodology.\

In the proposed scheme, in order to obtain the maximum expected long-term total discounted reward during a time period, the optimization goal is to determine which action should be taken. Let $J_{k}(\pi(k))$ be a value function, it represents the maximum expected reward that can be obtained from $\delta t_{k}$, given the belief state $\pi(k)$ at the beginning of time slot. Suppose that the MTCD that attempts to access the network makes action $a(k)$ and observes state $\theta(k)$, the reward can be accumulated starting from $\delta t_{k}$. The reward includes two parts: one is the immediate reward $Re_{l,n}$, the other is the maximum expected future reward $J_{k+1}(\pi(k+1))$ starting from $\delta t_{k+1}$, given the belief state $\pi(k+1)$~\cite{SY13}. Then, the value function is represented as
\begin{equation}\label{totalreward1}
\begin{split}
J_{k}(\pi(k))=\max\limits_{a(k)\in{\mathcal{A}}}\sum\limits_{s_r(k)\in{\mathcal{S}}}\sum\limits_{s_r(k+1)\in{\mathcal{S}}}\pi_{s_r(k)}^{k}p(s_r(k),s_r(k+1))\\ \sum\limits_{s_r(k)\in {\mathcal{S}}}b^{a(k)}_{s_r(k+1),\theta_r(k)}[Re_{l,n}(k)+J_{k+1}(\pi(k+1))],\\ \forall 1\leq k\leq K-1,
\end{split}
\end{equation}
and after a time period, the maximum expected reward is calculated as
\begin{equation}\label{totalreward2}
\begin{split}
J_{K}(\pi(K))=\max\limits_{a(k)\in{\mathcal{A}}}\sum\limits_{s_r(k)\in{\mathcal{S}}}\sum\limits_{s_r(k+1)\in{\mathcal{S}}} \pi_{s_r(k)}^{k}p(s_r(k),\\ s_r(k+1)) \sum\limits_{s_r(K)\in{\mathcal{S}}}b^{a(k)}_{s_r(K),\theta_r(K)}Re_{l,n}(K).
\end{split}
\end{equation}

According to Eqs. (\ref{totalreward1}) and (\ref{totalreward2}), it should be found that the optimization policy of random access with M2M communications will affect the total reward in two ways: on the one hand, how to obtain the immediate reward; on the other hand, how to transform and select the belief state that determines the future reward.\

\section{Simulation Results and Discussions}\label{sec:Simulation}
In this section, simulation results are presented to show the performance improvement of the proposed scheme modeled by POMDP. Network scenario is described in Figure~\ref{fig:model}. In order to ensure the comparison fairness, other schemes are listed as follows: the existing scheme without observation, random selection scheme and the existing scheme with perfect knowledge~\cite{ZY12}. For simplicity, the simulation is discussed by taking the highest class virtual vehicular network as an example.\

In the simulation, we assume a single-cell scenario with one eNodeB and 50 vehicles equipped with MTCDs. In addition, 25 RBs are offered by eNodeB. The physical vehicular network is sliced into 5 virtual networks based on different applications and QoS requirements. For each virtual vehicular network, one virtual eNodeB and multiple MTCDs are set. The number of MTCDs is set $30$ in the first virtual vehicular network, and the number of MTCDs in other virtual vehicular networks is set $5$, respectively. At beginning of each time slot, $5$ RBs allocate to each virtual eNodeB. Moreover, the probability that RB remains in the idle state, in the busy state, transits from busy to idle state or from idle to busy state is given as $0.85$, $0.1$, $0.9$ and $0.15$, respectively. Besides, $\rho$ and $\upsilon$ are assumed equal, they range from $0.1$ to $0.8$. Furthermore, the available transmission bandwidth in the first and other virtual networks are set $10$ MHz and $5$ MHz, respectively. The transmit power set $20$ dBm in each virtual vehicular network. Channel gains follow Gaussian distribution with zero mean and unit variance.\

\begin{figure}[!t]
\centering
\includegraphics[width=3.2in]{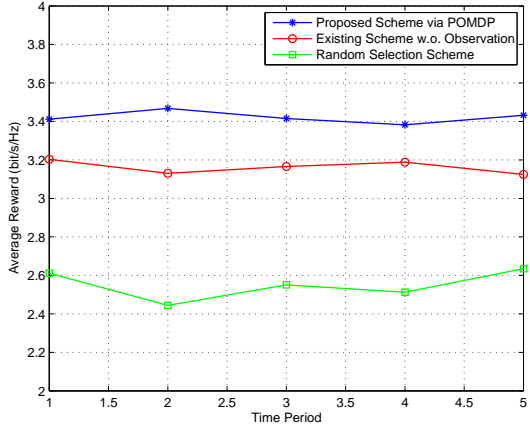}
\caption{Average reward within different time period in the heterogeneous traffic scenario.}
\label{fig:S1}
\end{figure}

As can be seen in Figure~\ref{fig:S1}, it illustrates the average reward improvement that is relative to the transmission capacity by the proposed scheme. Obviously, it can be seen that the proposed scheme has the higher reward than the existing scheme without observation and random selection scheme. Although the performance of the existing scheme without observation is closer to the proposed scheme, the advantages of the proposed scheme always exist in each the time period. According to POMDP optimization policy, the characteristic of observation can be shown by comparing with other algorithms. Moreover, the transmission capacity in the proposed scheme can be improved nearly $30\%$ compared with the random selection scheme. \

\begin{figure}[!t]
\centering
\includegraphics[width=3.2in]{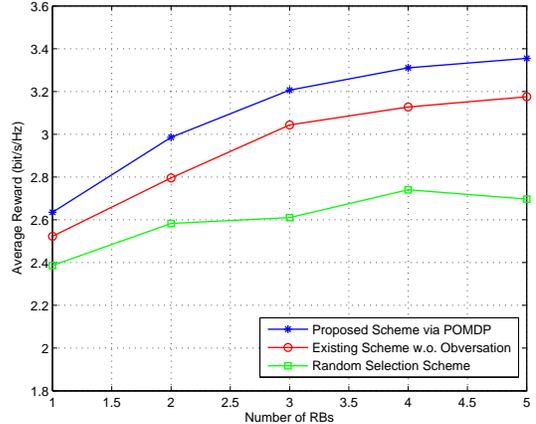}
\caption{Average reward with different number of RBs in the heterogeneous traffic scenario.}
\label{fig:S2}
\end{figure}

Figure~{\ref{fig:S2}} shows the reward achieved by the proposed scheme with those by other two existing schemes in different number of RBs. It easily can be found from Figure~{\ref{fig:S2}} that when the number of RBs is $1$, there is nearly no difference between these schemes. The reason is that the available number of RBs is less and there are not diverse decisions to make. However, with the increasing number of RBs, especially, when the number of RBs reaches $5$, the advantage of performance in the proposed scheme is prominent. Due to the increasing number of available RBs in the virtual network, more access chances are offered to MTCDs, then the received reward is also improved. Thus, more advantages will be shown if the proposed scheme applies into the network that has large number of RBs.\

\begin{figure}[!t]
\centering
\includegraphics[width=3.2in]{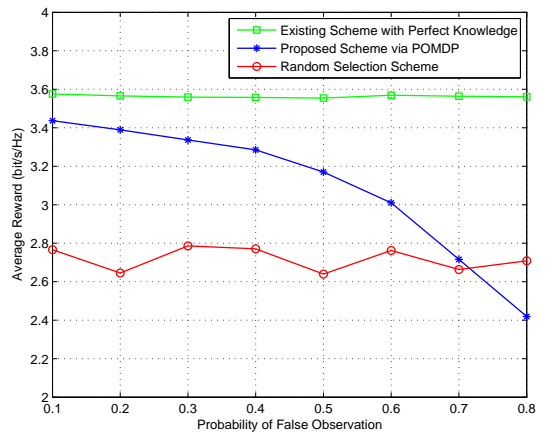}
\caption{Average reward with different probability of false observation in the heterogeneous traffic scenario.}
\label{fig:Feedback}
\end{figure}

Figure \ref{fig:Feedback} depicts the variation of average reward with different probability of false observation. The results show that the average reward degrades with the increasing probability of false observation. When the probability of false observation is $0.1$, the achieve reward in the proposed scheme will be close to the existing scheme with perfect knowledge. However, when the probability of false observation is $0.8$, the performance in the proposed scheme has degraded obviously. Especially, the performance is even worse than the random selection scheme when the probability of false observation is high. The reason is that the MTCDs have to give up or falsely select RBs to access when the probability of false observation is high. As a consequence, there is lower average reward to be obtained.\

\section{Conclusions and Future Work}\label{sec:Conclusion}
In this paper, we proposed a novel framework for M2M communications in vehicular networks with WNV. In the proposed framework, the physical resources are virtualized as virtual resources according to vehicles' functions and QoS requirements. As a consequences, the virtual eNodeB and the vehicles equipped with MTCDs that have the same or similar functions can form a virtual vehicular network. In addition, we formulated the random access process of vehicular as a POMDP, by which MTCDs can select proper RBs to achieve the maximum transmission rate. Simulation results were presented to demonstrate that, with the proposed framework,  the transmission rate achieved by MTCDs can be improved significantly. Future work is in progress to consider energy consumption and cooperative communications for vehicular networks in our framework.\

\section*{Acknowledgment}

This work is jointly supported by the National Natural Science Foundation of China under Grant No. 61372089 and No. 61571021.

\balance
\bibliographystyle{unsrt}
\bibliography{limengreferenceDIVANet,D:/CA/Papers/Ref}
\end{document}